\documentclass[preprint,aps,prb]{revtex4}

\usepackage{graphicx}
\begin{document}
\title{Orbital Ordering Structures in (Nd,Pr)$_{0.5}$Sr$_{0.5}$MnO$_3$ Manganite Thin Films on Perovskite (011) Substrates}
\author{Y.~Wakabayashi$^1$, D.~Bizen$^2$, Y.~Kubo$^3$, H.~Nakao$^2$,  Y.~Murakami$^{2}$, M.~Nakamura$^4$\footnote{Present address: Correlated Electron Research Center (CERC), National Institute of Advanced Industrial Science and Technology (AIST), Tsukuba, Ibaraki 305-8562, Japan}, Y.~Ogimoto$^5$,  K.~Miyano$^{3,6}$, and H.~Sawa$^1$}
\address{$^1$Photon Factory, Institute of Materials Structure Science, High Energy Accelerator Research Organization, Tsukuba 305-0801, Japan\\
$^2$Department of Physics, Tohoku University, Sendai 980-8578, Japan\\
$^3$Research Center for Advanced Science and Technology, University of Tokyo, Tokyo 153-8904, Japan\\
$^4$Department of Applied Physics, University of Tokyo, Tokyo 113-8586, Japan\\
$^5$Devices Technology Research Laboratories, SHARP Corporation, Nara 632-8567, Japan\\
$^6$CREST, JST, Kawaguchi 332-0012, Japan
}

\date{\today}
\begin{abstract}
Structural study of orbital-ordered manganite thin films has been conducted using synchrotron radiation, and a ground state electronic phase diagram is made. The lattice parameters of four manganite thin films, Nd$_{0.5}$Sr$_{0.5}$MnO$_3$ (NSMO) or Pr$_{0.5}$Sr$_{0.5}$MnO$_3$ (PSMO) on (011) surfaces of SrTiO$_3$ (STO) or [(LaAlO$_3$)$_{0.3}$(SrAl$_{0.5}$Ta$_{0.5}$O$_3$)$_{0.7}$] (LSAT), were measured as a function of temperature. The result shows, as expected based on previous knowledge of bulk materials, that the films' resistivity is closely related to their structures. Observed superlattice reflections indicate that NSMO thin films have an antiferro-orbital-ordered phase as their low-temperature phase while PSMO film on LSAT has a ferro-orbital-ordered phase, and that on STO has no orbital-ordered phase. A metallic ground state was observed only in films having a narrow region of {\it A}-site ion radius, while larger ions favor ferro-orbital-ordered structure and smaller ions stabilize antiferro-orbital-ordered structure. The key to the orbital-ordering transition in (011) film is found to be the in-plane displacement along [$0\bar 1 1$] direction.
\end{abstract}

\maketitle

\section{Introduction}
Perovskite manganese oxides have attracted considerable interest because of their various electronic states given by the complex combination of spin, charge, orbital, and lattice degrees of freedom.\cite{Tokukra00Science} The idea of orbital physics, which treats the relation between the orbital degree of freedom and the material property, influences a wide range of solid state physics\cite{Pen97PRL,Paolasini99PRL,Mizokawa99PRB,Khomskii05PRL}. Since the orbitals in manganites are strongly coupled with the lattice, structural studies are vital to elucidate the physical properties of manganites. 

The thin film technique is a method for applying huge biaxial pressure to material samples. Since anisotropic pressure has a dominant influence on the orbital state, this technique is useful to study the orbital state in manganites with a perturbation through the lattice. Reference [\onlinecite{Konishi99JPSJ}] reports that the biaxial strain caused by perovskite (001) substrates allows control of the ferro-orbital order (ferro-OO) structure in manganites. This example shows the intimate coupling between the orbital and the lattice. The strong coupling that allows control of the ferro-OO state, however, prohibits realizing antiferro-OO manganite films or films having a first-order metal-insulator transition that involves a change in symmetry.\cite{Prellier00PRB,Ogimoto01APL,Biswas01PRB,Buzin01APL} Thin films having a clear metal-insulator transition have recently been obtained by changing the orientation of the substrate from (001) to (011),\cite{Ogimoto05PRB,Nakamura05APL} and our previous report\cite{Wakabayashi06PRL} showed that one of such films has an antiferro-OO state. Manganite thin film on a (011) substrate is not a carbon copy of bulk manganite, but carries a new physics of its own, such as new type of colossal magnetoresistance\cite{Uozu06PRL} and a photo-induced reversible phase transition\cite{Takubo05PRL}. Microscopic study such as diffraction measurement is indispensable to clarifying the key factors of the system's unique properties.

In this paper, the results of a structural study of four manganite thin films, Nd$_{0.5}$Sr$_{0.5}$MnO$_3$ (NSMO) or Pr$_{0.5}$Sr$_{0.5}$MnO$_3$ (PSMO) on (011) surfaces of SrTiO$_3$ (STO, $a=3.905$~\AA) or [(LaAlO$_3$)$_{0.3}$(SrAl$_{0.5}$Ta$_{0.5}$O$_3$)$_{0.7}$] (LSAT, $a=3.870$~\AA) are reported. Using the structures obtained, the orbital arrangements of them are derived based on the assumption of a strong-coupling between the orbital and the lattice. Although recent $L$-edge RXS measurement shows\cite{HerreroMartin06PRB} an orbital arrangement that cannot be derived from the MnO$_6$ distortion, the mode of the distortion still provides a lot of information about the orbital as the $Q$-space that can be measured by $L$-edge RXS measurements is small. 

\begin{figure}
\includegraphics[width=7cm]{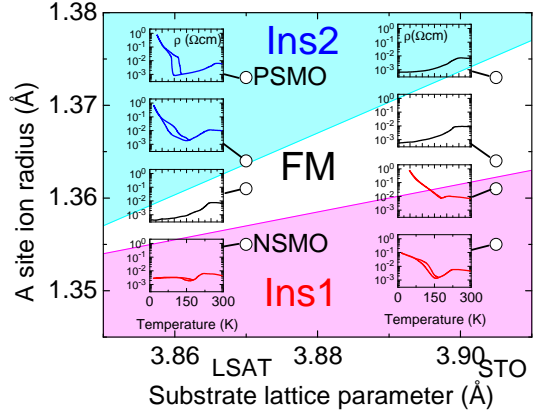}
\caption{(color online) Ground state phase diagram of half-doped manganite thin films on (011) substrates. The resistivities of the films as functions of temperature\protect{\cite{Ogimoto05PRB}} are also shown. There are two insulating phases, i.e., a phase with a small {\it A}-site ion (Ins1) and that with a large {\it A}-site ion (Ins2), and one ferromagnetic metal phase (FM).}
\label{fig:properties}
\end{figure}

The physical properties of bulk NSMO and PSMO was reported in refs.[\onlinecite{Kawano97PRL,Kajimoto99PRB}]. The CE-type antiferromagnetic phase and relevant orbital ordered phase (CE-OO) appears only in a narrow region of hole concentration around half-doping in NSMO system, while half-doped PSMO system shows A-type antiferromagnetic phase. We thus expect different electronic structures would be observed also in NSMO films and PSMO films. The properties of the four films together with the solid solution of NSMO and PSMO\cite{Ogimoto05PRB} and a tentative ground state phase diagram are presented in Fig.~\ref{fig:properties}. 
There are two insulating phases, a phase with a small {\it A}-site ion (Ins1) and that with a large {\it A}-site ion (Ins2), separated by a ferromagnetic metallic phase (FM). The result of our microscopic study clarifies the electronic structures of the two insulating phases. Moreover, it also shows how the strain affects the electronic states.

Here we summarize the properties of the films we measured.\cite{Ogimoto05PRB} All the films we measured have a ferromagnetic metal phase whose Curie temperatures are about 250K. 
PSMO/STO does not have a metal-insulator transition at lower temperature while the other films do. The transition temperature $T_{\mbox{\scriptsize  MI}}$ is about 170K for NSMO films, and 94K (126K) for PSMO/LSAT in a cooling (heating) run. Optical transmittance spectra, which can be translated into optical conductivity as a function of frequency, show that all the films have a 3$d$ electron lobe spread in the [100] direction.

\section{Experiment}
Epitaxial films were grown using the pulsed laser deposition method\cite{Nakamura05APL,Ogimoto05PRB}. The typical thickness of each of the samples was 80~nm. The electronic property of the films was characterized by measuring optical transmittance spectra in the energy region of 0.25eV to 0.8eV. Infrared (IR) light from an IR lamp was monochromatized with a grating and linearly polarized in the [0$\bar 1$1] direction. The transmittance of the film was obtained by dividing the transmitted intensity of the sample by that for a bare substrate. The difference in the reflectance between the two is small
and structureless so that it was ignored.

X-ray diffraction experiments were carried out on the BL-4C and the BL-16A2 at the Photon Factory, KEK, Japan and on the X22C at the National Synchrotron Light Source, Brookhaven National Laboratory. The beamlines are equipped with standard four-circle diffractometers connected to closed-cycle refrigerators. The temperature dependence of the lattice parameters was measured for all the samples with x-rays having energies of 16keV, 9.5keV, or 6.5keV. Superlattice reflections were searched with x-rays having those energies, and their energy spectra were also measured around the Nd, Pr, and Mn absorption edges.

\section{Results and analysis}
\subsection{Optical Transmission Spectra}
\begin{figure}
\includegraphics[width=7cm]{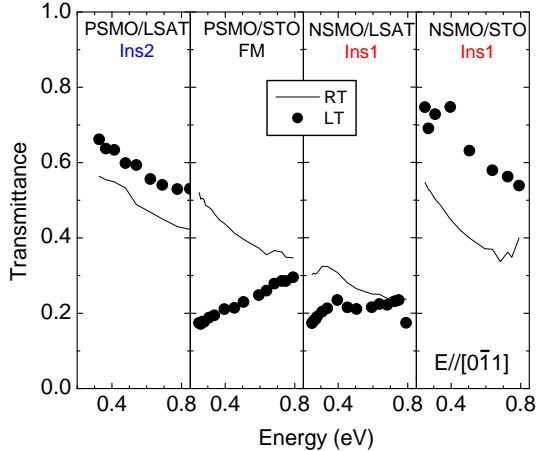}
\caption{Transmittance spectra of the films measured at room temperature and at low temperatures. All the spectra measured at room temperature show typical insulating features. At low temperatures, those of PSMO/LSAT and NSMO/STO show insulating feature and that of PSMO/STO shows metallic feature as expected from the DC resistivity. The transmittance spectrum for NSMO/LSAT, whose ground state is small {\it A}-site ion insulator (Ins1) with relatively small resistivity, shows a flat energy dependence, which is neither a typical metallic feature nor an insulating feature. }
\label{fig:spectrum}
\end{figure}

First, we characterized the electronic states of the samples by measuring IR transmittance spectra. Figure~\ref{fig:spectrum} shows the results measured at room temperature and those at low temperatures (lower than 30~K). At room temperature, all of the four spectra show insulating behavior, i.e., decreasing transmittance with increasing photon energy. Since all the films have a paramagnetic insulating phase at room temperature\cite{Ogimoto05PRB}, these spectra agree with the DC resistivity. At low temperatures, the spectrum for PSMO/STO has a metallic character (increasing transmittance with increasing photon energy) while those for NSMO/STO and PSMO/LSAT remain insulating; these optical properties are in close agreement with the DC resistivity shown in Fig.\ref{fig:properties}. On the other hand, the transmittance spectrum of NSMO/LSAT at low temperature does not show obvious insulating behavior, which implies that this sample includes both insulating and metallic phases. Based on the results of transmittance measurements, no significant difference between Ins1 and Ins2 was observed.

\subsection{Small {\it A}-site Ion Insulator Nd$_{0.5}$Sr$_{0.5}$MnO$_3$/SrTiO$_3$}
\begin{figure}
\includegraphics[width=7cm]{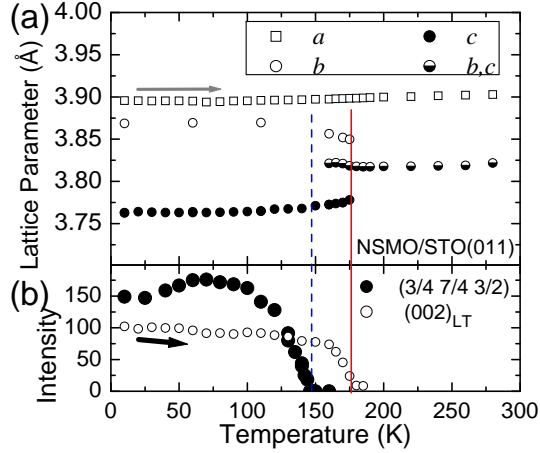}
\caption{(a) Temperature dependence of the lattice parameters $a$, $b$, and $c$ for NSMO/STO during a heating run measured with x-rays of 6.5~keV. The lattice parameter of SrTiO$_3$ is shown as $a$ because it is locked to the lattice parameter of the substrate which can be measured precisely. (b) Temperature dependence of the (002) Bragg reflection intensity of the low-temperature phase and that of the ($\frac34 \frac74 \frac32$) superlattice reflection intensity measured with x-rays of 6.552keV.}
\label{fig:LatCon_NSMO_STO}
\end{figure}

Figure~\ref{fig:LatCon_NSMO_STO} (a) shows the temperature dependence of the lattice parameters of NSMO/STO. Two-phase coexistence was observed around 170~K. Throughout this paper, lattice parameters are written in cubic notation because the Bravais lattice of the films is different from that of the bulk samples. The range of the ordinate for the lattice parameters of the four films is fixed at 3.7\AA\/ - 4.0\AA\/ for the sake of comparison. The lattice constants at room temperature are $a=3.905$~\AA, $b=c=3.824$~\AA, $\alpha=90.5^\circ$, and $\beta =\gamma =90.3^\circ$, and those at 10~K are $a=3.896$~\AA, $b=3.867$~\AA, $c=3.761$~\AA, $\alpha$=90.4$^\circ$, $\beta$=90.1$^\circ$, and $\gamma$=90.6$^\circ$. The lattice parameter $a$ is unchanged at $T_{\mbox{\scriptsize  MI}}=170$K and locked to the substrate, while that for [$0\bar11$] is free as reported earlier\cite{Ogimoto05PRB,Nakamura05APL}. Unlike the lattice parameter $a$, lattice parameters $b$ and $c$ vary 
drastically at $T_{\mbox{\scriptsize  MI}}$. Clearly, this change in lattice parameters is related to the metal-insulator transition.

We searched for the superlattice reflections in the wide reciprocal space region, and found superlattice reflections characterized by the modulation vectors $\mathbf q_m=$ ($\frac14 \frac14 0$), ($\frac14 \frac14 \frac12$), ($\frac12 \frac12 0$), and ($\frac12 \frac12 \frac12$). Note that we did not find the superlattice reflections having $\mathbf q_m=$ ($\frac14 \frac14 0$) in our previous study,\cite{Wakabayashi06PRL}  in which we measured only in the region of $0.5 \le h \le 1$ and $1.5\le \{k,l\} \le 2$; The superlattice reflection ($\frac34 \frac74 2$), the only superlattice position having $\mathbf q_m=$($\frac14 \frac14 0$) in this region, has accidentally too small structure factor to measure. 
Superlattice reflections having $\mathbf q_m=$($\frac12 \frac12 \frac12$) were observed at all the temperatures we measured, while the others were observed only at low temperatures. It should be noted that the bulk NSMO with GdFeO$_3$-type lattice modulation has reflections having $\mathbf q_m=$(00$\frac12$) and ($\frac12 \frac12$0), while they are {\it not} observed in the film at room temperature, as well as those having $\mathbf q_m=$($\frac12 \frac12 \frac12$). This implies that the mode of the lattice modulation of the film in the high-temperature phase is similar to La$_{0.6}$Sr$_{0.4}$MnO$_3$ film on (001)-surface of SrTiO$_3$\cite{Wakabayashi04PRB}. The intensity of the (002) Bragg reflection of the low-temperature phase and that of the ($\frac34 \frac74 \frac32$) superlattice reflection intensity as functions of temperature are shown in Fig.~\ref{fig:LatCon_NSMO_STO} (b). The $\sqrt 2 \times 2\sqrt 2\times 2$ superstructure disappears at 150~K ($=T_{\mbox{\scriptsize  AF-OO}}$) during the heating run. This transition temperature is significantly lower than $T_{\mbox{\scriptsize  MI}}$ at which the lattice parameters $b$ and $c$ split. 

\begin{figure}
\includegraphics[width=7cm]{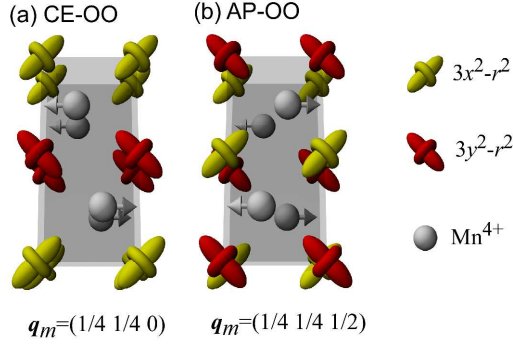}
\caption{(color online) Schematic view of the possible OO structures: (a) CE-type, and (b) AP(anti-phase)-type. Yellow, red, and white symbols represent $3x^2-r^2$, $3y^2-r^2$, and Mn$^{4+}$ (which means Jahn-Teller inactive ions), respectively. $\mathbf q_m$ for the orbital structure coincides with that for Mn$^{4+}$ displacements.}
\label{fig:OO_str_NSMO_STO}
\end{figure}

\begin{figure}
\includegraphics[width=7cm]{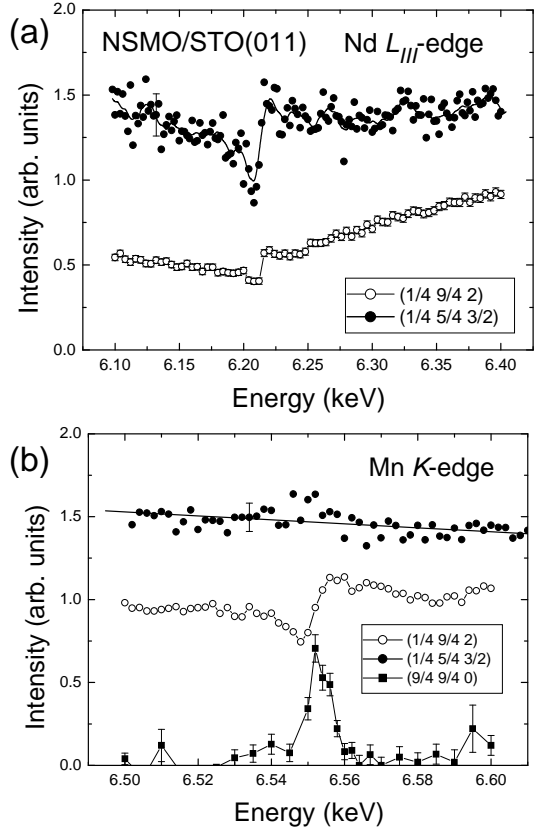}
\caption{Energy spectra of the superlattice reflections ($\frac14 \frac94 2$), ($\frac14 \frac54 \frac12$), and ($\frac94 \frac94 0$) measured around (a) Nd $L_{III}$-edge and (b) Mn $K$-edge. The reflections having modulation vector ($\frac14 \frac14 0$) contain amplitude from Mn ions, while that having modulation vector ($\frac14 \frac14 \frac12$) does not. }
\label{fig:RXS_NSMO_STO}
\end{figure}

Since our previous report analyzed the orbital order structure based on data with no intensity for $\mathbf q_m=$($\frac14 \frac14 0$), the structure must be reexamined. Take the two possible OO structures in this film shown in Fig.~\ref{fig:OO_str_NSMO_STO}, the CE-OO and anti-phase-OO (AP-OO) proposed in our previous paper\cite{Wakabayashi06PRL}, for example. Throughout this paper, we assume that the distortion of MnO$_6$ octahedra from the shape in the ferromagnetic metal phase, in which the orbital state is disordered, reflects the anisotropy of the $e_g$ electrons. The anisotropic otcahedra displace neighboring Mn ions as shown in the figure\cite{Radaelli97PRB}. Consequently, the wavevector of the Mn displacements in the transverse mode reflects that of the orbital order. In order to clarify the mode of Mn displacements, Mn $K$-edge RXS measurement, which is a useful method to obtain a lattice distortion around a specific element\cite{Wakabayashi04PRB}, was employed. Figure~\ref{fig:RXS_NSMO_STO} shows the energy spectra of several superlattice reflection intensities. 
A striking feature in these spectra is the lack of an intensity anomaly at the Mn $K$-edge in the spectrum of ($\frac14 \frac54 \frac12$) superlattice reflection, which means that the Mn ions do not have structural modulation having $\mathbf q_m=$($\frac14 \frac14 \frac12$). The ($\frac14 \frac54 \frac12$) superlattice reflection is mainly caused by the $A$-site displacement, which is indicated by the sharp kink at the Nd absorption edge shown in panel (a). In contrast, both the Mn and Nd sites have an atomic displacement having $\mathbf q_m=$($\frac14 \frac14 0$) because the spectrum of ($\frac14 \frac54 2$) shows sharp kink at both edges. The lattice modulation with $\mathbf q_m=$($\frac14 \frac14 0$) is a transverse mode modulation because of no superlattice reflections at ($h \pm \frac14, h \pm \frac14,0$) at non-resonant energies. The ($\frac14 \frac14 0$) transverse lattice modulation of Mn ions means that the OO structure has $\mathbf q_m=$($\frac14 \frac14 0$), so-called CE-type orbital ordering. In the course of this measurement, we found no superlattice reflection at ($\bar \frac94 \frac94 0$), ($\frac94 0 \frac94$) and ($\bar \frac94 0 \frac94$) while we found a resonant peak at ($\frac94 \frac94 0$), showing that only ($\frac14 \frac14 0$) is the wavevector of the lattice modulation caused by the orbital ordering.

Between $T_{\mbox{\scriptsize  MI}}$ and $T_{\mbox{\scriptsize  AF-OO}}$, the film has lattice parameters $a\simeq b > c$ without any superlattice reflections. The $a$, $b$ and $c$-lattice parameters are twice the Mn-O distances in three directions when we neglect the effect of the octahedra tilting, which is quadratic to the distance; It is justified by ref.[\onlinecite{Wakabayashi04PRB}], in which we made clear the oxygen positions in manganite films and found that the lattice parameter ratio qualitatively represents the Mn-O bond length ratio. Therefore a MnO$_6$ octahedron of NSMO/STO in the intermediate temperature phase has two long Mn-O bonds and one short bond. Since we assume the distortion of the octahedron reflects the anisotropy of the $e_g$ electrons, the orbital state in this phase is found to be a ferro-OO with an $x^2-y^2$-orbital, i.e., A-type ferro-OO (A-OO). Note that the lattice parameters for A-OO and those for CE-OO in bulk manganites are very similar\cite{Kajimoto99PRB}, which shows a good agreement with this case.

\subsection{Small {\it A}-site Ion Insulator Nd$_{0.5}$Sr$_{0.5}$MnO$_3$/LSAT}
\begin{figure}
\includegraphics[width=7cm]{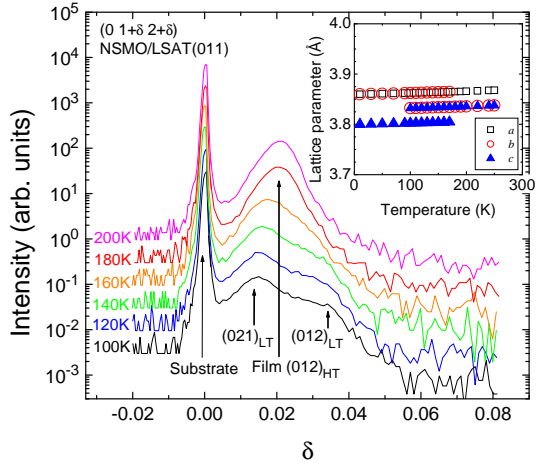}
\caption{(color online) The (012) peak profiles of NSMO/LSAT along the [011] direction at several temperatures. (inset) Temperature dependence of the lattice parameters measured with x-rays of 9.5~keV. }
\label{fig:LatCon_NSMO_LSAT}
\end{figure}

\begin{figure}
\includegraphics[width=7cm]{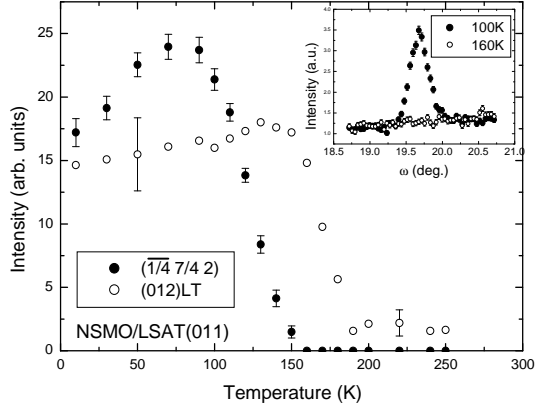}
\caption{Temperature dependence of the integrated intensity of ($\bar \frac14 \frac 74 2$) superlattice reflection (closed circle) and (012) Bragg reflection for the low-temperature phase (open circle) measured with x-rays of 9.5~keV. The finite intensity of the latter above 200K is caused by the tail of (012) reflection for the high-temperature phase. (inset) Peak profile of ($\bar \frac14 \frac 74 2$) superlattice reflection.}
\label{fig:Super_NSMO_LSAT}
\end{figure}

Figure~\ref{fig:LatCon_NSMO_LSAT} shows the (012) peak profile along [011]-direction at several temperatures. The (012) peak for the high-temperature phase ((012)$_{\mbox{\scriptsize HT}}$) clearly splits into two peaks ((012)$_{\mbox{\scriptsize LT}}$ and (021)$_{\mbox{\scriptsize LT}}$) at low temperatures. The peak intensity of the film Bragg reflection having a large [0$\bar 1$1] component in low-temperature phase was very weak. This forced us to analyze the lattice parameters based on the (012) reflection profile. Since the Debye-Waller factor is written by the average of $\exp(-\vec Q \cdot \vec r)$, where $\vec Q$ and $\vec r$ denote the scattering vector and atomic displacement vector\cite{Guinier}, this intensity decreasing in the [0$\bar 1$1] direction can be interpreted as  a large fluctuation in atomic displacement along this direction. As the peaks are too close to fit without any constraints, we analyzed the data with three-Gaussian curves at fixed peak widths and peak center positions for (012)$_{\mbox{\scriptsize HT}}$, (012)$_{\mbox{\scriptsize LT}}$ and (021)$_{\mbox{\scriptsize LT}}$. The temperature dependence of the (012) Bragg reflection intensity in low-temperature phase in a heating run was obtained through this three-Gaussian fitting, and the result is shown in Fig.~\ref{fig:Super_NSMO_LSAT}. The intensity of the (012)$_{\mbox{\scriptsize LT}}$ reflection decreased between 150K and 170K=$T_{\mbox{\scriptsize  MI}}$, and the peaks corresponding to the high-temperature phase (not shown) increases significantly in a heating from 150K to $T_{\mbox{\scriptsize  MI}}$.  The temperature dependence of the lattice parameters is shown in the inset of Fig.~\ref{fig:LatCon_NSMO_LSAT}; the fixed peak center gives no temperature variation of the lattice parameters except for the temperature dependence of the lattice parameter of the substrate. The lattice parameter $a$ is fixed to that of the substrate, and the value of lattice parameter $b$ in the low-temperature phase coincides with that of $a$. We also found an excess Bragg reflection around the (013) position caused by another minor phase that emerges below 190K. The intensity ratio of this peak to the (013) Bragg reflection of the high temperature phase was less than 20\%. The lattice in this phase is released from the substrate. When the sample is warmed up above 200K, the released phase disappears and merges into the major phase.

Superlattice reflection ($\bar \frac14 \frac 74 2$) was measured as a function of temperature. Although the peak intensity is much weaker than that in NSMO/STO, we observed the finite intensity of this reflection. The inset of Fig.~\ref{fig:Super_NSMO_LSAT} shows the peak profile, while the main panel shows the temperature dependence of the integrated intensity. Judging from the superlattice reflection with $\mathbf q_m=$($\frac14 \frac14 0$), this film also has a CE-OO state below $T_{\mbox{\scriptsize  AF-OO}}$=150K in its main phase. The intermediate state between $T_{\mbox{\scriptsize  MI}}$ and $T_{\mbox{\scriptsize  AF-OO}}$ has the lattice parameters $a \sim b > c$, and following the same discussion with previous section, it is again the A-OO state.

As a result, the structure of NSMO/LSAT film was found to be very similar to that of NSMO/STO film except for the large atomic positional fluctuation in the [0$\bar 1$1] direction and the existence of the minor phase. Although the semi-metallic behavior of the optical transmittance may be caused by the coexistence of the minor phase, there is another possibility that the major phase has such intermediate property. The mechanism is discussed in section \ref{sec:NSMO_LSAT}.

\subsection{Metallic Film Pr$_{0.5}$Sr$_{0.5}$MnO$_3$/SrTiO$_3$}
\begin{figure}
\includegraphics[width=7cm]{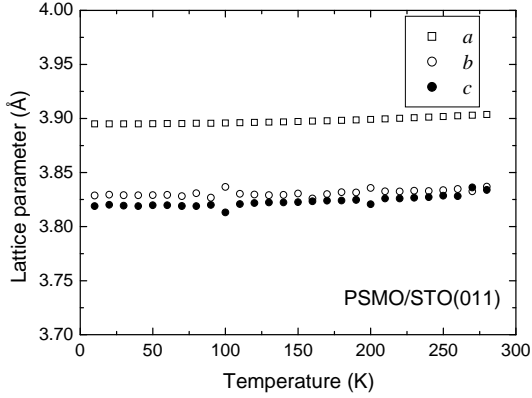}
\caption{Temperature dependence of the lattice parameters of PSMO/STO(011) measured with x-rays of 16~keV. }
\label{fig:LatCon_PSMO_STO}
\end{figure}

As can be seen in Fig.~\ref{fig:properties} and ref.[\onlinecite{Ogimoto05PRB}], this film has no low-temperature insulating phase. Figure~\ref{fig:LatCon_PSMO_STO} shows the temperature dependence of the lattice parameters of this film. 
No anomaly can be seen in this figure. We also searched for the superlattice reflections, and no superstructure was found. This result shows that the entire film was in the metallic phase down to 10K, and no different phases coexisted.

\subsection{Large {\it A}-site Ion Insulator Pr$_{0.5}$Sr$_{0.5}$MnO$_3$/LSAT}
\begin{figure}
\includegraphics[width=7cm]{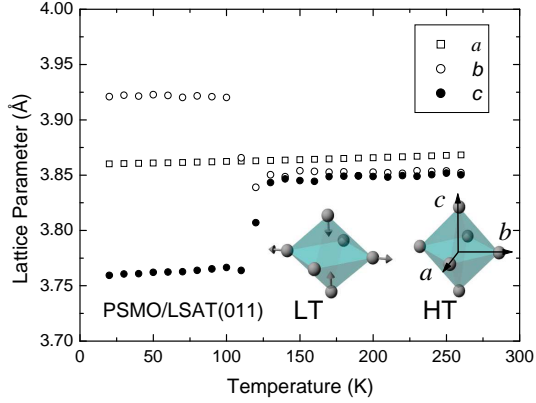}
\caption{(Color online) Temperature dependence of the lattice parameters of PSMO/LSAT(011) in a heating run measured with x-rays of 6.5~keV. (inset) Schematic view of MnO$_6$ octahedra in the high-temperature phase and the low-temperature phase. }
\label{fig:LatCon_PSMO_LSAT}
\end{figure}

This film has an insulating phase below 126K (94K) in a heating (cooling) run.\cite{Ogimoto05PRB} Figure~\ref{fig:LatCon_PSMO_LSAT} shows the temperature dependence of the lattice parameters in a heating run. The transition temperature is between 120K and 130K, which corresponds to the metal insulator transition. The jump in the lattice parameters on a cooling run happens at 30K lower than heating run, the same width of the hysteresis observed in the resistivity measurements. Therefore, the lattice parameter splitting is accompanied by the insulating phase and no phase-coexistence happens in this film.
The lattice parameters $a$, $b$, and $c$ are very close to each other in the high-temperature phase. In the low-temperature phase, the three parameters differ widely from each other. 

We found no superlattice reflections at 20K at positions characterized by the wavevectors \{$\frac 14 \frac14 0$\} and \{$\frac 14 \frac14 \frac12$\}, or by the modulation vector in bulk Pr$_{0.5}$Sr$_{0.5}$MnO$_3$\cite{Kajimoto02PRB}, \{$0, 0, 0.3$\}. This result clearly shows that the insulating phase of this film has different atomic/electronic structures than those of NSMO films. Judging from the change in lattice parameters and absence of the superstructure, the low-temperature insulating phase is a $\cos \frac{\theta}{2} |3z^2-r^2\rangle+\sin \frac{\theta}{2} |x^2-y^2\rangle$ ($2\pi/3 \le \theta \le \pi$) ferro-OO phase, which is an intermediate orbital between $x^2-y^2$ and $3y^2-r^2$ orbitals, which favor A-type and C-type antiferro magnetic structures, respectively (In this paper, we call it A/C ferro-OO). The distortion mode of the low-temperature phase is shown in the inset to Fig. \ref{fig:LatCon_PSMO_LSAT} together with that of the high-temperature phase. The low-temperature phase has the lattice parameters of $b>a>c$, which means that the $e_g$ orbital has long lobe along $b$-direction, short lobe along $a$-direction and no lobe along $c$-direction. The linear combination of the two $e_g$ orbitals with $2\pi/3 \le \theta \le \pi$ reproduces this character. Note that the orbital states in the intermediate temperature phases of NSMO films, which were classified into A-OO, have little difference in $a$ and $b$ lattice parameters.

\section{Discussion}
\subsection{Free Energy of Distorted Lattice --- Difference Between (001)-film and (011)-film}\label{sec:NSMO_LSAT}
\begin{figure}
\includegraphics[width=7cm]{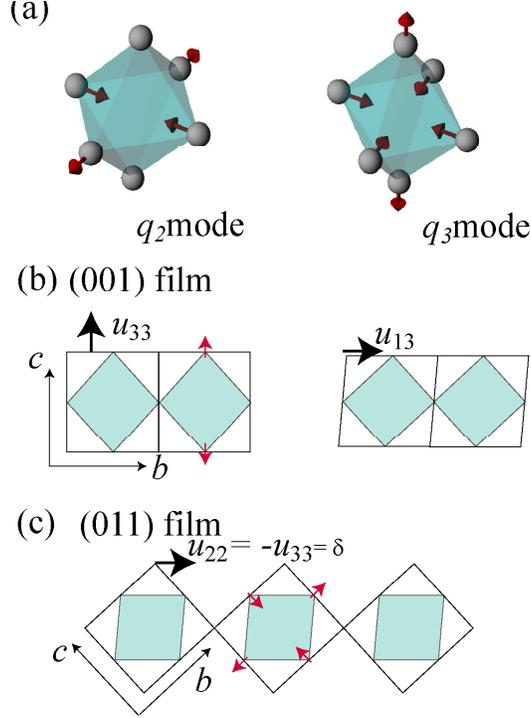}
\caption{(Color online) (a) Two JT modes with $e_g$ symmetry, $q_2$ and $q_3$. (b) Schematic view of the lattice strain in (001) film. The JT mode strain $u_{33}$ causes the volume change, while the non-JT mode $u_{13}$ does not.  (c) That for (011) film. The shear-mode strain $u_{22}=-u_{33}=\delta$ is the JT-mode $q_2$. }
\label{fig:strain}
\end{figure}

It has been pointed out that a clear first-order phase transition accompanied by structural change is demonstrated in films only when they are formed on (011) substrates \cite{Ogimoto05PRB,Nakamura05APL}. It is instructive to discuss the reason why this is the case before proceeding with a discussion of the structure.
In the bulk NSMO, the changes in the lattice constants at the ferromagnetic transition are less than 0.2\%\cite{Kawano97PRL}. In contrast, those at the CE-OO transition amount to 1.4\%, while the volumetric change is 0.1\% \cite{Kawano97PRL,Shimomura99JPSJ}. The volumetric change $\delta V=\sum_i u_{ii}$ ($u_{ij}$: strain tensor fixed to the cube axes) obviously requires very much energy and is not favored under any circumstances. Especially in the film system, the volumetric change at the transition temperature was less than the experimental error, 0.03\%. This is because of the large value of the diagonal element of the strain tensor $\sigma_{ii}$ caused by the substrate and the thermodynamical relation $\partial F/\partial u_{ij}=\sigma_{ij}$ where $F$ denotes the free energy. Therefore, in this section, we assume the $\delta V$ is forced to zero. Meanwhile, due to the Jahn-Teller (JT) instability of 1/4 filled $e_g$ orbitals, particular types of lattice distortion can happen spontaneously, the so-called JT distortion modes ($q_2$ and $q_3$ modes in Fig. \ref{fig:strain}). The collective JT distortion is the driving force for the charge- and orbital-ordering. 

For a film grown on a (001) substrate, the constraint $u_{11}=u_{22}=0$ automatically imposes $u_{33}=0$. The distortion of the type $u_{13}$ is not prevented but the $q_2$ component is second-order in $u_{13}$  under the condition $u_{11}=u_{22}=u_{33}=0$ and does not couple to the JT mode. On the other hand, for a film grown on a (011) substrate, the deformation constraints due to the substrate are $u_{11}=0$ and $u_{23}=0$, where the former is the constraint from the substrate itself, and the latter is derived from the constraint $u_{22}+u_{33}-2u_{23}=0$ and $\delta V=u_{11}+u_{22}+u_{33}=0$. The shear deformation $u_{22}=-u_{33}=\delta$ ($\delta$ denotes a finite value) is allowed because the value of $\delta V$ caused by this strain is zero, and the $q_2$ mode is linear in this distortion. Thus, the transition from the isotropic metallic phase to the charge- and orbital-ordered phase can proceed in the film without any barrier once the collective JT distortion has lower energy. The shear mode deformation in the [0$\bar 1$1] direction changes the system from metallic to charge-ordered. This mode of structural fluctuation was found in NSMO/LSAT which has both a CE-OO structure and semi-metallic transmittance; the structural fluctuation perhaps makes the system partially metallic.

Many phase transitions to the CE-OO state in the bulk are dominated by the $q_2$ mode, hence the good coincidence between the bulk and film MI transition. However, it should be commented that the $q_3$ mode is not allowed in films and this constraint brings about a phase sequence unique to thin films as shown in the next section.

\subsection{Phase Diagram}
\begin{figure}
\includegraphics[width=7cm]{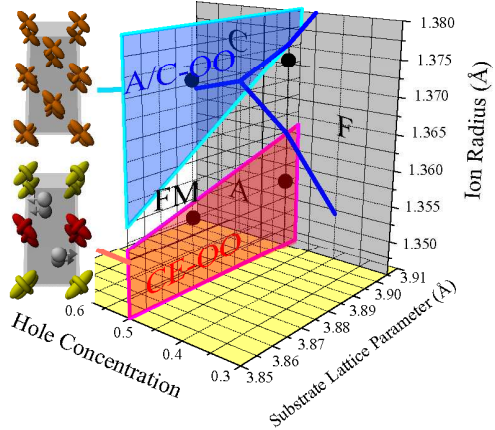}
\caption{(color online) Ground state phase diagram of manganite films on a (011) substrate derived by our study together with a schematic ground state diagram of manganite films given by theoretical calculation\protect{\cite{Konishi99JPSJ}}. The theoretical phase diagram was derived in hole concentration versus lattice parameter ratio $c/a$, and in this figure, $c/a$ was converted to the ion radius for presentation. }
\label{fig:Phase_diagram}
\end{figure}

The results show that the NSMO films have CE-OO, PSMO/STO has metallic, and PSMO/LSAT has A/C ferro-OO phases as their ground states. 
 It is clear that the tolerance factor alone, which is widely used for bulk manganites, is not sufficient to characterize the thin film because this factor is only a function of the {\it A}-site ion radius while the electronic states are also influenced by the lattice parameters of the substrate.
We summarized our data as a ground state phase diagram in the {\it A}-site ion radius vs the substrate lattice parameter plane; it is shown in Fig.~\ref{fig:Phase_diagram}.  {\it A}-site ions of a certain size give little lattice distortion and ferromagnetic metal phase. Smaller ions make the lattice parameters $a\simeq b\gg c$ resulting in A-OO and CE-OO, while larger ions make lattice parameters $b>a>c$ resulting in A/C ferro-OO. The phase diagram differs qualitatively from that for bulk manganite which does not have such a wide intermediate metallic phase around the half-doped region. Therefore, one may expect that the phase diagram having intermediate metallic phase is caused by an anisotropic pressure; we examined this idea with comparing the phase diagram with that for the films on (001) substrates\cite{Konishi99JPSJ} which is dominated by the anisotropic pressure.

The calculated phase diagram for thin films on a (001) substrate is shown in Fig.~\ref{fig:Phase_diagram} with gray shade, in which, for the sake of illustration, the variable $c/a$ in the original word is translated into the {\it A}-site ion radius. For $x=0.5$, this phase diagram shows the intermediate metallic phase together with small-ion A-OO phase and large-ion C-OO phase. The similarity is more prominent when we recall that the CE-OO phase has A-OO as its higher temperature phase and that A/C ferro-OO has some C-OO components in the present study. Therefore, the strain from the substrate, which prohibits volumetric change as well as $q_3$ mode distortion, plays an important role in determining the ground state phase diagram for films on (011) substrates, while the high temperature electronic structures in them have larger freedom than the films on (001) substrates.

The small {\it A}-site ion radius favors the CE-OO phase. Extrapolating this tendency, we expect a CE-OO phase for the low-temperature phase of the manganite film with much smaller {\it A}-site ions Pr$_{0.55}$(CaSr)$_{0.45}$MnO$_3$/LSAT(011), which is reported\cite{Takubo05PRL} to have a persistent photoinduced phase transition. Our preliminary measurements show a CE-OO phase in these films, supporting this result. Although the phase boundary is not identical, it was found  qualitatively that the smaller {\it A}-site ions favor CE-OO and larger ones favor metallic phase in the Pr$_{0.55}$(CaSr)$_{0.45}$MnO$_3$/LSAT(011), the same trend with present study.

\section{Conclusion}
We have for the first time derived the ground state phase diagram of half-doped manganite thin films on (011) substrates with microscopic probe. The results show that the system has two kinds of insulating phases: One is an A-type or CE-OO phase having a two-dimensional nature with lattice parameters $a\simeq b\gg c$, and the other is an A/C-OO phase having a one-dimensional feature with lattice parameters $b>a>c$. As in the phase diagram of films on (001) substrates, the large effect of the strain on the films' electronic states is elucidated. The insulating phase discussed in ref.[\onlinecite{Uozu06PRL}], an insulating behavior ascribed to the low-dimensionality, was classified as an A/C ferro-OO phase with a more one-dimensional character.  Moreover, the mode that allows the orbital ordering transition was found to be the in-plane atomic displacement along [$0\bar1 1$] direction.

\section*{Acknowledgements}
The authors are grateful to Prof. T.~Arima and Dr. J.~P.~Hill for their fruitful discussions. This work was supported by a Grant-in-Aid for Creative Scientific Research (13NP0201) and TOKUTEI (16076207) from the Ministry of Education, Culture, Sports, Science and Technology of Japan and JSPS KAKENHI (15104006). A part of this work was conducted under support for long-term visits from the Yamada Science Foundation. Financial support to M. N. by the 21st Century COE Program for ``Applied Physics on Strong Correlation'' administered by the Department of Applied Physics, The University of Tokyo is also appreciated. Use of the National Synchrotron Light Source, Brookhaven National Laboratory, was supported by the U.S. Department of Energy, Office of Science, Office of Basic Energy Sciences, under Contract No. DE-AC02-98CH10886.


\begin{thebibliography}{99}

\bibitem{Tokukra00Science} Y.~Tokura and N.~Nagaosa, Science {\bf 288} 462 (2000).

\bibitem{Pen97PRL} H.~F.~Pen, J.~van~den~Brink, D.~I.~Khomskii and G.~A.~Sawatzky: Phys. Rev. Lett. {\bf 78}, 1323 (1997).

\bibitem{Paolasini99PRL} L.~Paolasini, C.~Vettier, F.~de~Bergevin, F.~Yakhou, D.~Mannix, A.~Stunault, W.~Neubeck, M.~Altarelli, M.~Fabrizio, P.~A.~Metcalf, and J.~M.~Honig, Phys. Rev. Lett. {\bf 82}, 4719 (1999).

\bibitem{Mizokawa99PRB} T.~Mizokawa, D.~I.~Khomskii, and G.~A.~Sawatzky, Phys. Rev. B. {\bf 60}, 7309 (1999).

\bibitem{Khomskii05PRL} D.~I.~Khomskii and T.~Mizokawa, Phys. Rev. Lett. {\bf 94}, 156402 (2005).

\bibitem{Konishi99JPSJ} Y.~Konishi, Z.~Fang, M.~Izumi, T.~Manako, M.~Kasai, H.~Kuwahara, M.~Kawasaki, K.~Terakura and Y.~Tokura, J. Phys. Soc. Jpn. {\bf 68}, 3790, (1999).


\bibitem{Prellier00PRB} W.~Prellier, Ch.~Simon, A.~M.~Haghiri-Gosnet, B.~Mercey, and B.~Raveau, Phys. Rev. B {\bf 62}, R16337, (2000).

\bibitem{Ogimoto01APL} Y.~Ogimoto, M.~Izumi, T.~Manako, T.~Kimura, Y.~Tomioka, M.~Kawasaki, and Y.~Tokura, Appl. Phys. Lett. {\bf 78}, 3505, (2001).

\bibitem{Biswas01PRB} A.~Biswas, M.~Rajeswari, R.~C.~Srivastava, T.~Venkatesan, R.~L.~Greene, Q.~Lu, A.~L.~de~Lozanne, and A.~J.~Millis, Phys. Rev. B {\bf 63}, 184424 (2001).

\bibitem{Buzin01APL} E.~R.~Buzin, W.~Prellier, Ch.~Simon, S.~Mercone, B.~Mercey, B.~Raveau, J.~Sebek, and J.~Hejtmanek, Appl. Phys. Lett. {\bf 79}, 647 (2001).

\bibitem{Ogimoto05PRB} Y.~Ogimoto, M.~Nakamura, N.~Takubo, H.~Tamaru, M.~Izumi, and K.~Miyano, Phys. Rev. B {\bf 71}, 060403(R), (2005).

\bibitem{Nakamura05APL} M.~Nakamura, Y.~Ogimoto, H.~Tamaru, M.~Izumi, and K.~Miyano, Appl. Phys. Lett. {\bf 86}, 182504 (2005).

\bibitem{Wakabayashi06PRL} Y.~Wakabayashi, D.~Bizen, H.~Nakao, Y.~Murakami, M.~Nakamura, Y.~Ogimoto, K.~Miyano and H.~Sawa, Phys. Rev. Lett. {\bf 96}, 017202 (2006).

\bibitem{Uozu06PRL} Y.~Uozu, Y.~Wakabayashi, Y.~Ogimoto, N.~Takubo, H.~Tamaru, N.~Nagaosa, and K.~Miyano, Phys. Rev. Lett. {\bf 97}, 037202 (2006).

\bibitem{Takubo05PRL} N.~Takubo, Y.~Ogimoto, M.~Nakamura, H.~Tamaru, M.~Izumi, and K.~Miyano, Phys. Rev. Lett. {\bf 95}, 017404 (2005).

\bibitem{HerreroMartin06PRB}	J.~Herrero-Martin, J.~Garcia, G.~Subias, J.~Blasco, M.C.~Sanchez and S.~Stanescu, Phys. Rev. B 73, 224407 (2006).


\bibitem{Kawano97PRL} H.~Kawano, R.~Kajimoto, H.~Yoshizawa, Y.~Tomioka, H.~Kuwahara and Y.~Tokura, Phys. Rev. Lett. {\bf 78} 4253 (1997).

\bibitem{Kajimoto99PRB} R.~Kajimoto, H.~Yoshizawa, H.~Kawano, H.~Kuwahara, Y.~Tokura, K.~Ohoyama, and M.~Ohashi, Phys. Rev. B {\bf 60} 9506 (1999).


\bibitem{Wakabayashi04PRB} Y.~Wakabayashi, H.~Sawa, M.~Nakamura, M.~Izumi, and K.~Miyano, Phys. Rev. B  {\bf 69} 144414 (2004).


\bibitem{Radaelli97PRB} P.~G.~Radaelli, D.~E.~Cox, M.~Marezio, and S-W.~Cheong, Phys. Rev. B {\bf 55}, 3015 (1997).


\bibitem{Guinier} A.Guinier, X-ray diffraction, W.H.Freeman and co., San Francisco (1963).



\bibitem{Kajimoto02PRB} R.~Kajimoto, H.~Yoshizawa, Y.~Tomioka, and Y.~Tokura, Phys. Rev. B {\bf 66}, 180402(R) (2002).

\bibitem{Shimomura99JPSJ} S.~Shimomura, K.~Tajima, N.~Wakabayashi, S.~Kobayashi, H.~Kuwahara, and Y.~Tokura, J. Phys. Soc. Jpn. {\bf 68} 1943 (1999).


\end{thebibliography}
\end{document}